\documentclass[sigconf]{acmart}
\AtBeginDocument{%
  }

\copyrightyear{2025}
\acmYear{2025}
\setcopyright{cc}
\setcctype{by}
\acmConference[SIGIR '25]{Proceedings of the 48th International ACM SIGIR Conference on Research and Development in Information Retrieval}{July 13--18, 2025}{Padua, Italy}
\acmBooktitle{Proceedings of the 48th International ACM SIGIR Conference on Research and Development in Information Retrieval (SIGIR '25), July 13--18, 2025, Padua, Italy}\acmDOI{10.1145/3726302.3730189}
\acmISBN{979-8-4007-1592-1/2025/07}

\usepackage{acronym}
\usepackage{balance}

\begin{document}

\acrodef{IR}[IR]{Information Retrieval}
\acrodef{IIR}[IIR]{Interactive Information Retrieval}
\acrodef{LLM}[LLM]{Large Language Model}
\acrodefplural{LLM}[LLMs]{Large Language Models}
\acrodef{sDCG}[sDCG]{Session-based DCG}
\acrodef{idf}[idf]{inverse document frequency}
\acrodef{CSM}[CSM]{Complex Searcher Model}
\acrodef{SERP}[SERP]{search engine result page}
\acrodef{DCG}[DCG]{discounted cumulated gain}
\acrodef{RBP}[RBP]{rank-biased precision}
\acrodef{sRBP}[sRBP]{Session RBP}
\acrodef{IG}[IG]{information gain}
\acrodef{SERP}[SERP]{search engine result page}
\acrodefplural{SERP}[SERPs]{search engine result pages}


\title{Evaluating Contrastive Feedback for Effective User Simulations}
\author{Andreas Konstantin Kruff}
\authornote{These authors contributed equally to this research.}
\orcid{0009-0002-8350-154X}
\affiliation{
  \institution{TH Köln}
  \city{Cologne}
  \country{Germany}
}
\email{Andreas.Kruff@th-koeln.de}

\author{Timo Breuer}
\authornotemark[1]
\orcid{0000-0002-1765-2449}
\affiliation{
  \institution{TH Köln}
  \city{Cologne}
  \country{Germany}
}
\email{Timo.Breuer@th-koeln.de}

\author{Philipp Schaer}
\orcid{0000-0002-8817-4632}
\affiliation{
  \institution{TH Köln}
  \city{Cologne}
  \country{Germany}
}
\email{Philipp.Schaer@th-koeln.de}

\renewcommand{\shortauthors}{Andreas Kruff, Timo Breuer, and Philipp Schaer}


\begin{abstract}
The use of Large Language Models (LLMs) for simulating user behavior in the domain of Interactive Information Retrieval has recently gained significant popularity. However, their application and capabilities remain highly debated and understudied. This study explores whether the underlying principles of contrastive training techniques, which have been effective for fine-tuning LLMs, can also be applied beneficially in the area of prompt engineering for user simulations.

Previous research has shown that LLMs possess comprehensive world knowledge, which can be leveraged to provide accurate estimates of relevant documents. This study attempts to simulate a knowledge state by enhancing the model with additional implicit contextual information gained during the simulation. This approach enables the model to refine the scope of desired documents further. The primary objective of this study is to analyze how different modalities of contextual information influence the effectiveness of user simulations. 

Various user configurations were tested, where models are provided with summaries of already judged relevant, irrelevant, or both types of documents in a contrastive manner. The focus of this study is the assessment of the impact of the prompting techniques on the simulated user agent performance. We hereby lay the foundations for leveraging LLMs as part of more realistic simulated users.

\end{abstract}


\begin{CCSXML}
<ccs2012>
   <concept>
       <concept_id>10002951.10003317.10003331</concept_id>
       <concept_desc>Information systems~Users and interactive retrieval</concept_desc>
       <concept_significance>300</concept_significance>
       </concept>
 </ccs2012>
\end{CCSXML}

\ccsdesc[300]{Information systems~Users and interactive retrieval}

\keywords{Interactive Information Retrieval (IIR), SimIIR 3, Contrastive Feedback, User Simulation}


\maketitle

\section{Introduction}

The aim of this work is to evaluate the search effectiveness of LLM-based user agents by comparing different prompting strategies to simulate a human-like knowledge state. We hypothesize that, as demonstrated in other fine-tuning contexts, contrastive learning may enable the LLM to internalize task-specific distinctions more effectively, resulting in improved interaction capabilities compared to other prompting strategies presented here. \acp{LLM} have shown a remarkable influence in current \ac{IR} research. While some of their applications, like the ability of \acp{LLM} regarding relevance judgments, are highly debated and controversial, other use cases, like LLM-based query reformulations, have shown their potential. 
Query reformulations can be used to increase retrieval performance~\cite{DBLP:conf/sigir-ap/000224}, and the effectiveness of LLM-based query reformulations has been shown, especially when the LLM was allowed to rely on itself or when provided with additional TREC topic descriptions~\cite{engelmann2024contextdriveninteractivequerysimulations}. Query reformulations are also a crucial step in modern user simulations~\cite{DBLP:journals/ftir/BalogZ24} and in related frameworks like SimIIR 3~\cite{DBLP:conf/sigir-ap/Azzopardi00KM0P24}. Therefore, while still understudied, \acp{LLM} show a high potential for user simulations. 

The knowledge state can be modeled by incorporating the information need that is typically described in a structured format given by the topic title, description, narrative, and furthermore, the relevance labels (qrels) of on \ac{IR} test collection. 
However, we argue that in a real-world scenario, the users are uncertain about their explicit information needs~\cite{belkinAnomalousStatesKnowledge1980}, and the realism of the user simulation when proving the \ac{LLM} with the explicitly formulated topic descriptions can be questioned. Instead, this study tries to iteratively update the simulated user's knowledge state by summaries of already seen and judged documents. The study evaluates whether supplying summaries of relevant documents, irrelevant documents, or a combination of both aid the model in decision-making and query formulation.

\section{Related Work}

Early attempts of user simulations in \ac{IIR} were characterized by the use of rule-based user agents. This includes, for example, the Common Interaction Model (CIM), which handles the different user interactions like query reformulation, relevance judgment or stopping behaviour by deterministc rules like stopping after a predefined number of irrelevant document~\cite{CIM}. The same applies for the Complex Searcher Model (CSM) which was used in the first version of the SimIIR framework \cite{SimIIR1,SearchingStopping}. While these approaches are transparent and allow easy interpretation, their effectiveness is limited, especially when trying to simulate real user behavior. Recent studies have shown the potential capabilities of \acp{LLM} regarding their use in language understanding and interactive decision making tasks, which combine reasoning, acting, and planning. It has been shown that these capabilities can be further improved by applying approaches like chain-of-thought prompting or action plan generation~\cite{ReAct,zhou2024languageagenttreesearch}.
These led to the recent development of extended frameworks like SimIIR 3~\cite{DBLP:conf/sigir-ap/Azzopardi00KM0P24} and USimAgent~\cite{USim} that allow the use \acp{LLM} for simulating user interactions. The effectiveness of LLM-based compared to advanced rule-based interactions remains understudied.

Recent studies have shown how contrastive learning approaches can improve the performance of \acp{LLM} in various ways. These include, among others, fewer hallucinations in conversational  usecases~\cite{sun2022contrastivelearningreduceshallucination}, improvements in few-shot text classification~\cite{zhang-etal-2024-la} or few-shot natural language understanding tasks~\cite{gunel2021supervisedcontrastivelearningpretrained}. These versatile improvements in fine-tuning \acp{LLM} by contrastive learning led to the hypothesis that these findings are also applicable and transferable for improvements in prompting strategies for user simulations.

Using \acp{LLM} for certain one-shot-learning tasks is a recent and highly debated topic in the community. The usage of \acp{LLM} for creating relevance judgments is of great importance to the \ac{IR} community due to its crucial role in the whole evaluation process~\cite{Soboroff_2025,10.1145/3673791.3698431,takehi2024llmassistedrelevanceassessmentsask}.  
However, this discussion is still ongoing and there are recent venues like the LLM4Eval SIGIR workshop that push the discussion forward with contributions like LLMJudge~\cite{rahmani2024llmjudgellmsrelevancejudgments}. We need to synthesize the relevance decision in our simulated \ac{IIR} in some way or the other. Therefore, we take advantage of the new approaches but would like to point to the ongoing discussion.

\section{Methodology}
\begin{figure*}[!ht]
\centering

\includegraphics[width=1\textwidth]{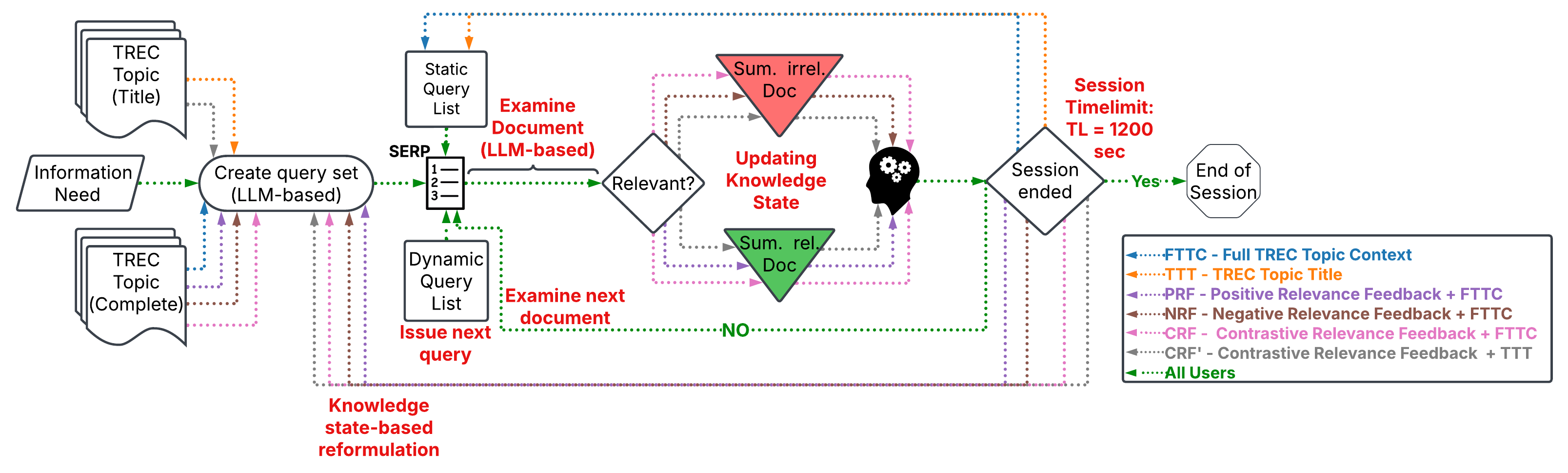}
\caption{A flowchart of the analyzed user configuration described in Section~\ref{subsec:user_config}. The color codes correspond to those in Figure~\ref{fig:experimental_results} that show the experimental results in the simulated search sessions.}
\label{fig:experimental_setup}
\end{figure*}

In this study, different prompting strategies are evaluated regarding their potential to improve the user agent's search effectiveness in simulated search sessions. In particular, we evaluate how the effectiveness of the LLM-based user agent changes if context information based on the already seen documents is added to pre-existing context information that represents the user's knowledge state. In our simulations, the context information is obtained from the topic files of test collections. TREC test collections provide information needs in topic files alongside the document collection and the corresponding relevance judgments. These topic files are formatted in a de-facto standardized format and usually cover a title, a more detailed description, and an additional narrative to provide annotators with context information when making relevance judgments. Our simulated users are instantiated with different combinations of the topic file's information sections by including them in a prompt given to an LLM. The language model then decides about (ir-)relevant documents that will eventually be added to the user's knowledge state throughout the simulation. Overall, our methodology is a multi-layered approach that covers the combination of available information items, i.e., user configurations  (cf.~\ref{subsec:user_config}), the prompting technique  (cf.~\ref{subsec:prompt_technique}), and the LLM's ability to generate effective queries and make correct decisions about the relevance of single results (cf.~\ref{subsec:dataset_implementation}). Ultimately, our experiments answer the following research question:

\begin{itemize}
    \item [RQ] What influence does the modality of context information in the prompt have on the effectiveness of LLM-based user simulations?
\end{itemize}

where our null hypothesis would therefore be that no differences between the different prompting strategies are to be expected. The entire experimental setup, including the code and results are accessible via our public GitHub repository.\footnote{\url{https://github.com/irgroup/SIGIR2025-Contrastive_Prompting}}

\subsection{User Configurations}
\label{subsec:user_config}

Our experiments cover three different types of user configurations: baseline users based on intuitive methods and heuristics, user configurations that either exploit relevant or irrelevant information items judged by the employed LLM, and another contrastive approach that makes use of both relevant and irrelevant information obtained throughout the progression of the simulation. Based on these three types, we define different variants or subtypes that lead to a total of eight different users that are evaluated in the experiments. In the following, we provide detailed information of how each particular user type and its variants are defined. Figure~\ref{fig:experimental_setup} provides an overview of what kinds of simulation steps are implemented by the different configurations, which we plan to complement in future work with LLM-based user agents for additional comparisons.

\textbf{Baseline Users:} These user configurations act as baselines and help to contextualize the results of other implemented user agents. For two of the baselines, we make use of the TREC topic file's contents, where the \textbf{Full TREC Topic Context (FTTC)} user includes the title, description and narrative for the query generation and relevance decisions, and the \textbf{TREC Topic Title (TTT)} is a derived variant that simply includes the topic's title. The queries for these approaches are generated only once at the beginning of the session by the LLM. In addition, we also include random relevance decision configurations to better contextualize the LLM's ability to make correct relevance judgments. The \textbf{random user (RND)} is based on a naive query generation approach that generates \textit{three-term-queries} based on the vocabulary given by the topic file. The relevance decision are made with a randomized Bayesian decision process and a probability $p=0.5$. In addition, we also evaluate another \textbf{random user with pre-determined queries (RND*)} that uses the same relevance decision approach but makes use of the LLM's generated queries of FTTC.

\textbf{Positive/Negative Relevance Feedback (PRF/NRF) :} These two user configurations have access to the same contextual information as the FTTC user in the beginning of the simulation and utilize the same prompt until the first document is deemed (ir)relevant. Once the first document is judged, a summary of all earlier seen documents is included in the prompt to assist the LLM with subsequent relevance decisions and query generation steps in the ongoing session. The main difference lies within the type of earlier seen documents. While the PRF user exclusively considers documents judged as relevant for the summarization, the NRF user only considers irrelevant documents.

\textbf{Contrastive Relevance Feedback (CRF):} This user configuration combines the above-mentioned approaches. When there are judgments for relevant or irrelevant documents, this user will be provided with both summaries. While the CRF use can access the entire topic's contents similar to FTTC, the CRF' is only provided with the topic's title and the summaries of both relevant and irrelevant documents seen earlier in the simulated session. This approach resembles the prevalent fine-tuning approach that includes both negative and positive training samples during training. The main difference lies in applying this paradigm during the downstream task, during the interactive prompting and simulation steps.

\subsection{Prompting Techniques}
\label{subsec:prompt_technique}

In the experiments, we use \textit{Few-shot Learning} (in-context learning) as this methodology aligns well with the interactive retrieval scenario and the idea that the user already has a certain not yet clearly defined information need. As explained before, this context can be conveyed either through the description and narrative provided by the topic files, through a summarization of previously viewed documents, or through a combination of both. Depending on the user configuration, the LLM receives a summarization of irrelevant documents, relevant documents, or a combination of both. The summarization was performed iteratively with each newly judged document, utilizing all previously encountered relevant and irrelevant documents according to the desired summary format. At the beginning of each prompt, the model was assigned a persona --- in this case, a journalist to align the model with the newswire corpora described in \ref{subsec:dataset_implementation}--- alongside a task-specific instruction. The prompting templates can be found in the provided code repository and an example prompt is provided in the README.\footnote{\url{https://github.com/irgroup/SIGIR2025-Contrastive_Prompting/blob/main/README.md}} 

\subsection{Implementation Details and Datasets}
\label{subsec:dataset_implementation}

The experiments were conducted on two TREC newswire test collections including the New York Times Annotated Corpus\footnote{\url{https://catalog.ldc.upenn.edu/LDC2008T19}} used as part of TREC Common Core 2017 (\textbf{Core17})~\cite{DBLP:conf/trec/AllanHKLGV17}, and the TREC Washington Post Corpus\footnote{\url{https://trec.nist.gov/data/wapost/}} used as part of TREC Common Core 2018 (\textbf{Core18})~\cite{DBLP:conf/trec/2018}. The experiments were implemented with the help of the newly introduced SimIIR 3 Framework~\cite{DBLP:conf/sigir-ap/Azzopardi00KM0P24}, with partially adapted logic. The newest iteration of the Framework provides a PyTerrier \cite{10.1145/3459637.3482013} integration, allowing for easy integration of existing TREC Collections. In the simulation experiments, BM25 was employed as the ranking function due to its efficiency, which is essential for interactive simulations with many query reformulations. While more effective ranking methods based on LLMs might retrieve better results, they come at the price of increased computational costs. In our study, the focus is on the simulated user interactions rather than the ranking method.

The Llama3.3 model in \texttt{Q4\_K\_M} quantization with 70B parameter was used for both the query generation and relevance judgment of the documents. While for the query generation step the temperature was set to $1.0$, allowing for more creativity in the of query generation process, the temperature was set to 0 for the relevance judgments in order to force the model to provide fully context-driven relevance decisions. We used a fixed random seed, with the default value set to $0.0$. All of the other parameters were also kept at the default configuration.

\subsection{Evaluation Measures}

Our experimental evaluation is twofold in order to validate the results from two different perspectives. First, we evaluate the trade-off between the \textbf{effort and effect} in a search session with the help of the \textit{information gain (IG)}. It is determined as follows:

\begin{equation}
    \mathrm{Effect} = \sum_{s \in \mathcal{S}} \mathrm{IG} (s), \qquad \mathrm{IG} (s) = \left\{\begin{array}{ll}
        \mathrm{rel}_d, & \mathrm{if} \ s = s_{\mathrm{rel}} \\ 
        0, & \mathrm{else}.
        \end{array} \right .
\end{equation}

where $s$ denotes a single logged interaction in a set of all logged interactions $\mathcal{S}$, and $s_{\mathrm{rel}}$ denotes the logged interaction that consider the item as relevant, contributing to the information gain.

Secondly, we evaluate the \textbf{Session-Discounted Cumulative Gain (sDCG)}~\cite{DBLP:conf/ecir/JarvelinPDN08} to align our contributions with earlier work. The main difference between these measures lies with the costs they consider, while the information gain is determined by all interaction costs including the query formulation, snippet inspection, document selection, relevance judgments etc., sDCG is more abstract by solely considering the query costs.

\section{Experimental Results}

\begin{figure}
    \centering
    \includegraphics[width=0.5\columnwidth]{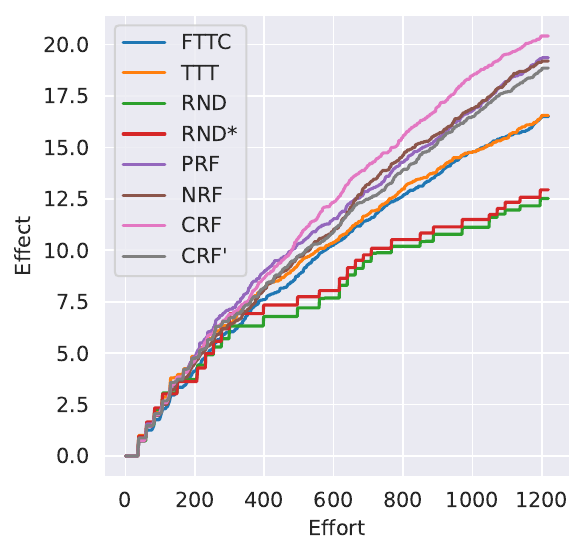}
    \includegraphics[width=0.475\columnwidth]{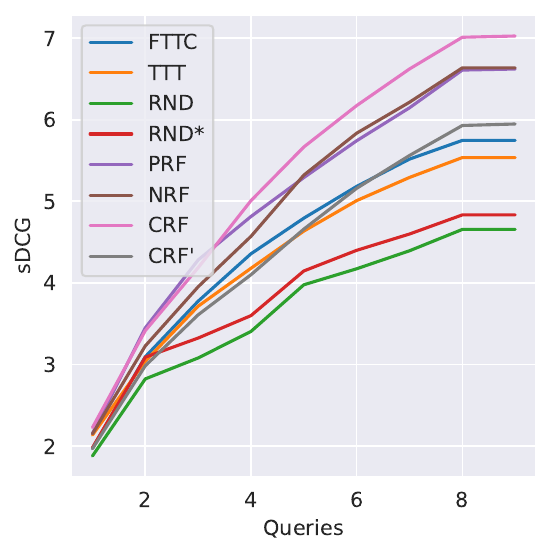}
    \includegraphics[width=0.5\columnwidth]{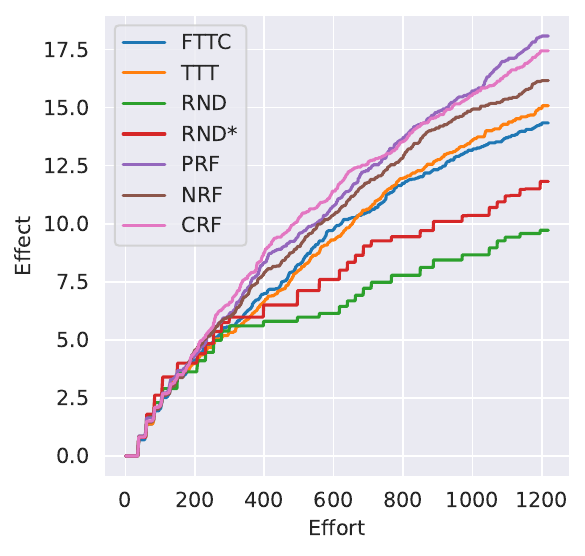}
    \includegraphics[width=0.475\columnwidth]{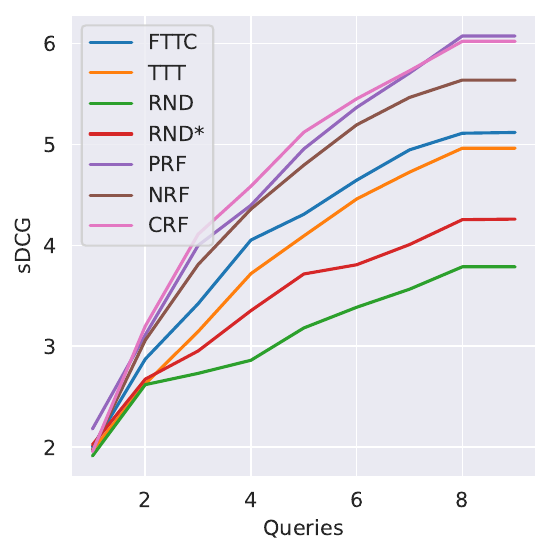}
    \caption{Results of the simulated sessions for different prompting strategies evaluated with the help of Effort-Effect trade-offs and sDCG on Core17 (top) and Core18 (bottom).}
    \label{fig:experimental_results}
\end{figure}

Figure~\ref{fig:experimental_results} shows the experimental results of our simulations. All figures show increasing scores as the simulated sessions progress, which indicates that our experimental setup is a viable environment for user simulations in general. Referring back to our research question, we can indeed identify differences between the different user configurations. In the early phase of the simulations, there are no apparent differences between the configurations. However, in the long run, these differences become more obvious, suggesting that there is an \textit{uptake} phase for the reliable use of the different configurations, e.g., for simulating synthetic interaction data.  

For all measures, the random user configuration \textbf{RND} and \textbf{RND*} serve as lower bounds for effectiveness within the constraints of our experimental setup. Overall, it is evident that the simulated users benefit from the LLM's relevance decisions. The \textbf{TTT} and \textbf{FTTC} users are more effective simulators than the naive baselines with random relevance decisions. While these users rely on different amounts of topic-related information, there is only a small and almost negligible difference between the search effectiveness, which highlights the potential use of LLMs in future simulation experiments, even with test collections that do not have complete topic files with descriptions and narratives like MS MARCO~\cite{DBLP:conf/sigir/CraswellMYCL21}.

Including the relevance information by seen documents in the sessions increases the search effectiveness even more, as suggested by the results of the \textbf{NRF}, \textbf{PRF}, \textbf{CRF}, and \textbf{CRF'} users. They surpass the TTT and FTTC users for all measures and both test collections. It can be observed that their effectiveness surpasses that of the other approaches --- at the latest after four queries have been issued. 

On Core17, the \textbf{CRF} user stands out as the top performer, highlighting the effectiveness of contrastive examples in the prompt. However, it is important to note that is essential to include the full topic information in this case, since there are apparent differences between \textbf{CRF} and \textbf{CRF'} --- the user configuration without any other topic-related information besides the title. This circumstance suggests that topical drifts might be an issue, and the topic's title might not be enough to give the LLM proper guidance.

On Core18, \textbf{CRF} is on par with the \textbf{PRF} user, which remains the leading performer overall with this test collection. Even though the differences are marginal, there is obviously no benefit to including summaries of irrelevant documents in the prompt in this case. Therefore, future research needs to investigate when \textbf{CRF} is a user simulator that performs reliably better. A first starting point could be to compare the statistics and distributions of relevance judgments, which are also an integral part of our simulations.

\section{Conclusion and Future Work}

Referring back to our research question, we can answer it with a positive primary outcome. Our experimental results suggest that, as hypothesized, LLMs benefit from providing contrastive examples in the prompt throughout interactive retrieval simulations. This finding is further supported by the observation that even without using topic descriptions and narratives, the simulated users with contrastive examples performed comparably well to those users who considered the full topic information (including the description and narrative), as well as positive document examples seen in the simulated session.

Furthermore, our simulations reveal the limitations of ad hoc retrieval test collection, as some of the generated queries by the LLM retrieved documents that were not assessed for by the annotators during pooling. This circumstance underlines the need for additional and more elaborated resources to evaluate and validate simulated users of interactive search sessions.

A secondary outcome of this study was observed during our inspection of the generated interaction logs. For many queries, documents were retrieved without a relevance labels, which is typical for experiments with query variants. These query variants were not part of the pooling when the test collection was curated, and thus, some documents were simply not considered for relevance judgments. As a result, the underlying approaches could not benefit from finding additional potentially relevant documents without associated relevance judgments. Given that current work demonstrates the feasibility of using LLMs for reliable relevance judgments, future work may address the evaluation of simulations with unjudged documents that are considered relevant by an LLM. While this approach may help address the limitations of unjudged documents, we are aware that it may also introduce new challenges, as discussed in \cite{Soboroff_2025}. Currently, the community lacks any kind of efficient methodology to automatically evaluate these scenarios. There are also no resources for these kinds of simulations with extensive relevance evaluations, revitalizing the idea of having a test collection with complete judgments.

\begin{acks}
This work is partially funded by Deutsche Forschungsgemeinschaft (DFG) under grant number 509543643 and within the funding programme FH-Personal (PLan CV, reference number 03FHP109) by the German Federal Ministry of Education and Research (BMBF) and Joint Science Conference (GWK).

\end{acks}

\balance
\bibliographystyle{ACM-Reference-Format}
\bibliography{references}

\end{document}